\documentclass[11pt]{article}

\let\epsilon=\varepsilon

\usepackage{fullpage}
\usepackage{epsfig}
\usepackage{amssymb}

\newtheorem{thm}{Theorem}
\newtheorem{lem}[thm]{Lemma}

\newtheorem{cor}[thm]{Corollary}

\newtheorem{claim}[thm]{Claim}

\newenvironment{pf}{{\bf Proof:}}{\hfill $\Box$ \bigbreak}

\begin{document}

\title{The Lazy Bureaucrat Scheduling Problem\thanks{
Appears in preliminary form in
{\em Proceedings of the Workshop on Discrete Algorithms (WADS)},
Vancouver, BC, Canada, Aug 11--14, 1999, pages 
122--133~\cite{ArkinBenderMitchellSkiena99}.}}

\author{Esther M. Arkin\thanks{estie@ams.sunysb.edu. 
Department of Applied Mathematics and Statistics.
State University of New York, Stony Brook, NY 11794-3600.
Partially supported by 
the National Science Foundation (CCR-9732220, CCR-0098172) and
grants from ISX Corp.\ and HRL Laboratories.}
\and
Michael A. Bender\thanks{bender@cs.sunysb.edu.
Department of Computer Science.
State University of New York, Stony Brook, NY 11794-4400.
Partially supported by 
grants from ISX Corp., HRL Laboratories, Sandia National 
Labs, and the National Science Foundation (EIA-0112849, CCR-0208670).}
\and
Joseph S. B. Mitchell\thanks{jsbm@ams.sunysb.edu.
Department of Applied Mathematics and Statistics.
State University of New York, Stony Brook, NY 11794-3600.
Supported in part by 
HRL Laboratories, Honda Fundamental Research Labs, ISX Corp., Metron Aviation, Inc., Sandia National 
Labs, Seagull Technologies, Sun Microsystems, and 
the National Science Foundation (CCR-9732220, CCR-0098172).}
\and
Steven S. Skiena\thanks{skiena@cs.sunysb.edu.
Department of Computer Science.
State University of New York, Stony Brook, NY 11794-4400.
Supported in part by
the National Science Foundation (CCR-9988112) and
by the Office of Naval Research (N00149710589).}
}

\maketitle

\begin{abstract}
  We introduce a new class of scheduling problems in which the
  optimization is performed by the worker (single ``machine'') who
  performs the tasks.  A typical worker's objective is to minimize the
  amount of work he does (he is ``lazy''), or more generally, to
  schedule as inefficiently (in some sense) as possible.  The worker
  is subject to the constraint that he must be busy when there is work
  that he {\em can\/} do; we make this notion precise both in the
  preemptive and nonpreemptive settings.  The resulting class of
  ``perverse'' scheduling problems, which we denote ``Lazy Bureaucrat
  Problems,'' gives rise to a rich set of new questions that explore
  the distinction between maximization and minimization in computing
  optimal schedules.
\end{abstract}


\section{Introduction}

Scheduling problems have been studied extensively from the point of
view of the objectives of the enterprise that stands to gain from the
completion of a set of jobs.  We take a new look at the problem from
the point of view of the workers who perform the tasks that earn the
company its profits.  In fact, it is natural to expect that some
employees may lack the motivation to perform at their peak levels of
efficiency, either because they have no stake in the company's profits
or because they are simply lazy.  

The following example illustrates
the situation facing a ``typical'' office worker, who may be one small
cog in a large bureaucracy:

\begin{quote}
{\em Example.}\quad It is 3:00~p.m., and Dilbert goes home at
5:00~p.m.  Dilbert has two tasks that have been given to him: one
requires 10 minutes, the other requires an hour.  If there is a task
in his ``in-box,'' Dilbert must work on it, or risk getting fired.
However, if he has multiple tasks, Dilbert has the freedom to choose
which one to do first.  He also knows that at~3:15, another task will
appear --- a 45-minute personnel meeting.  If Dilbert begins the
10-minute task first, he will be free to attend the personnel meeting
at~3:15 and then work on the hour-long task from~4:00 until~5:00.  On
the other hand, if Dilbert is part way into the hour-long job at~3:15,
he may be excused from the meeting.  After finishing the 10-minute job
by~4:10, he will have 50~minutes to twiddle his thumbs, iron his tie,
or enjoy engaging in other mindless trivia.  Naturally, Dilbert
prefers this latter option.
\end{quote}

An historical example of a situation where it proved crucial to
schedule tasks inefficiently is documented in the book/movie {\em
  Schindler's List\/}~\cite{Keneally-82}.  It was essential for the
workers and management of Schindler's factory to appear to be busy at
all times in order for the factory to stay in operation, but they
simultaneously sought to minimize their contribution to the German war
effort.

These examples illustrate a general and natural type of scheduling
problem, which we term the ``Lazy Bureaucrat Problem'' (LBP); the goal
of the LBP is to schedule jobs as {\em inefficiently\/} (in some
sense) as possible.  There exists a vast literature on scheduling; see
e.g., some of the recent
surveys~\cite{KargerSteinWein:1997,LLKS-93,Pinedo-95}.  The LBP
studies these traditional problems ``in reverse.''  Several other
combinatorial optimization problems have also been studied in reverse,
leading, e.g., to maximum
TSP~\cite{BarvinokJWW1998,Fekete1999,HassinRubinstein1998,KosarajuParkStein1994},
maximum cut~\cite{GoemansWilliamson1995}, and longest
path~\cite{KargerMotwaniRamkumar1997}; such inquiries often lead to a
better understanding of the structure and algorithmic complexity of
the original optimization problem.

\subsection{The Model} 
\label{subsec-model}

In this paper we schedule a set of jobs $1 \ldots n$ having processing
times (lengths) $t_1 \ldots t_n$ respectively.  Job $i$ {\em
  arrives\/} at time $a_i$ and has its {\em deadline\/} at time $d_i$.
We assume throughout this paper that $t_i$, $a_i$, and $d_i$ have
nonnegative integral values.  The jobs have {\em hard deadlines,\/}
meaning that each job $i$ can only be executed during its allowed
interval $I_i = [a_i, d_i]$; we also call $I_i$ the job's {\em
  window\/}. We let $c_i=d_i-t_i$ denote the {\em critical time\/} of
job~$i$; job~$i$ must be started by time $c_i$ if there is any chance
of completing it on time.

The jobs are executed on a single processor, the (lazy) {\em
bureaucrat\/}.  The bureaucrat executes only one job at a time.  

\subsubsection{Busy Requirement}

The bureaucrat chooses a subset of jobs to execute.  Since the
bureaucrat's goal is to minimize his effort, he would prefer to remain
idle all the time and to leave all the jobs unexecuted.  However, this
scenario is forbidden by what we call the {\em busy requirement,\/}
which stipulates that the bureaucrat work on an {\em executable\/}
job, if any executable jobs exist.  A job is {\em executable\/} if the
constraints of the LBP allow the job to be run.  Thus, in the
nonpreemptive setting a job $j $ is executable as long as it begun
during the interval $[a_j, c_j]$, at which point it is run to
completion.  In the preemptive setting the constraints that govern
whether or not a job can be executed are more complicated; see
Section~\ref{sec:preempt}.

\subsubsection{Objective Functions}

In traditional scheduling problems, if it is impossible to complete
the set of all jobs by their deadlines, one typically optimizes
according to some objective, e.g., to maximize a weighted sum of
on-time jobs, to minimize the maximum lateness of the jobs, or to
minimize the number of late jobs.  For the LBP we consider three
different objective functions, which naturally arise from the
bureaucrat's goal of inefficiency:
\begin{enumerate}

\item
{\em Minimize the total amount of time spent working\/} ---
This objective naturally appeals to a ``lazy'' bureaucrat.

\item
{\em Minimize the weighted sum of completed jobs\/} ---  In this paper
we usually assume that the weight of job $i$ is its length, $t_i$;
however, other weights (e.g., unit weights) are also of interest.
This objective appeals to a ``spiteful'' bureaucrat whose goal it is
to minimize the fees that the company collects on the basis of his
labors, assuming that the fee (in proportion to the task length, or a
fixed fee per task) is collected only for those tasks that are
actually completed.

\item
{\em Minimize the {\em makespan}, the maximum completion
time of the jobs\/} ---  This objective appeals to an ``impatient''
bureaucrat, whose goal it is to go home as early as possible, at the
completion of the last job he is able to complete.  He cares about the
number of hours spent at the office, not the number of hours spent
doing work (productive or otherwise) at the office.

Note that, in contrast with standard scheduling problems on one
processor, the makespan in the LBP varies; 
it is a function of which jobs have passed their
deadlines and can no longer be executed.

\end{enumerate}

\subsubsection{Additional Parameters of the Model}
As with most scheduling problems, additional parameters of the model
must be set.  For example, we must explicitly allow or forbid {\em
preemption\/} of jobs.  When a job is {\em preempted\/}, 
it is interrupted and may be resumed later at no additional cost.  
If we forbid preemption, then once a job is begun, 
it must be completed without interruptions.

We must also specify whether scheduling occurs {\em on-line\/} or
{\em off-line.\/} A scheduling algorithm is 
off-line
if all the jobs are known to the scheduler at the outset; it is
on-line if the jobs are known to the scheduler only as they arrive.
In this paper we restrict ourselves to off-line scheduling; we leave
the on-line case as an 
open problem.

\begin{figure*}[bt] 
\def\RESULT#1{\vbox{\vskip 2pt\hbox{#1}\vskip 0.5pt}}
\begin{center}
\begin{tabular}{|l||c||c|}
\hline
{\bf Instance} & {\bf Metric} & {\bf Complexity} \\
\hline \hline
$\bullet$ Unit-length jobs: $t_1 =  \cdots = t_n = 1$  & \RESULT{1}
& \RESULT{Polynomial time} \\
\hline
$\bullet$ Short job intervals:  \mbox{$\forall i \, \, d_i - a_i < 2 t_i $} & 
\RESULT{1-3}  &
\RESULT{Pseudo-polynomial time} \\
\hline
$\bullet$ Ratios $R = O(1)$  and $\Delta = O(1)$ & 
\RESULT{1-3}  &
\RESULT{Pseudo-polynomial time} \\
\hline
$\bullet$ Same arrival times: $a_1 = \cdots = a_n$ & \RESULT{1-3}  
&
\RESULT{Pseudo-polynomial time} \\
\RESULT{~~~~}  & \RESULT{~~~~}  & 
 \RESULT{(Weakly) NP-complete} \\
\RESULT{~~~~} & 
\RESULT{~~~~~~} & 
\RESULT{Hard to approximate} \\
\hline
$\bullet$ Job size ratio $\Delta = O(1)$
& \RESULT{1-3}
&
 \RESULT{Strongly NP-complete} \\
\RESULT{~~~~}  & \RESULT{~~~~}  &
 \RESULT{Hard to approximate} \\
\hline
$\bullet$ General LBP
& \RESULT{1-3}
&
 \RESULT{Strongly NP-complete} \\
\RESULT{~~~~}  & \RESULT{~~~~}  &
 \RESULT{Hard to approximate} \\
\hline
\end{tabular}
\end{center}
\caption{
  Summary of our nonpreemptive results; see Section~\ref{nonpreempt}.}
\label{tab:nonpreemtive-results}
\end{figure*}

\subsection{Our Results}

In this paper, we introduce the Lazy Bureaucrat Problem and develop
algorithms and hardness results for several versions of the LBP.  {From}
these results, we derive some general characteristics of this new
class of scheduling problems and describe (1) situations in which
traditional scheduling algorithms extend to the LBP and (2)
situations in which these algorithms no longer apply.

\subsubsection{No Preemption}
We prove that the LBP is NP-complete, as is often the case for
traditional scheduling problems. Thus, we focus on special cases to
study exact algorithms.  When all
jobs have unit size, optimal schedules can be found in polynomial
time.
The following three cases have pseudo-polynomial algorithms:
(1) when each job $i$'s interval $I_i$ is less 
than twice the processing time of job $i$;
(2) when the ratios of interval length to job length and longest job
to shortest job are both bounded; and (3) when all jobs arrive in the
system at the same time.
These last scheduling problems are solved using dynamic programming
both for Lazy Bureaucrat and traditional metrics.  Thus, in these
settings, the Lazy Bureaucrat metrics and traditional metrics are
solved using similar techniques.

{From} the point of view of approximation, however, the standard and
Lazy Bureaucrat metrics behave differently.  Standard metrics
typically allow polynomial-time algorithms having good approximation
ratios, whereas we show that the Lazy Bureaucrat metrics are difficult
to approximate.  This hardness derives more from the busy requirement
and less from the particular metric in question, that is the busy
requirement appears to render the problem substantially more
difficult.  (Ironically, even in standard optimization problems, the
management often imposes this requirement because it intuitively
appears desirable.)

\begin{figure*}[bt] 
\def\RESULT#1{\vbox{\vskip 2pt\hbox{#1}\vskip 0.5pt}}
\begin{center}
\begin{tabular}{|c|c||l|}
\hline
{\bf Preemption} & {\bf Metrics} & {\bf Complexity}\\
\hline \hline
\RESULT{I} & \RESULT{1-3} &
\RESULT{Polynomial Time} \\
\hline
\RESULT{II} & \RESULT{1-3} &
\RESULT{(Weakly) NP-complete,  even when $a_0 = \cdots = a_n$} \\
\hline
\RESULT{III} & \RESULT{1-3} &
\RESULT{(Weakly) NP-complete,  even when \mbox{$a_0 = \cdots = a_n$}} \\
\RESULT{~~~} & \RESULT{~~~} & \RESULT{Hard to approximate} \\
\hline
\end{tabular}
\end{center}
\caption{Summary of our preemptive results; see Section~\ref{sec:preempt}.}
\label{tab:preemtive-results}
\end{figure*}

\subsubsection{Preemption}

The busy requirement dictates that the worker must stay busy while
work is in the system. If the model allows preemption we must specify
under what conditions a job can be interrupted or resumed. We
distinguish three versions of the preemption rules, which we list from
most permissive to most restrictive. In particular,
workers are constrained to execute the following jobs:
\begin{description}
\item[] (I) any job that has arrived and is before its deadline,
\item[] (II) any job that has arrived and for which there is still
  time to complete it before its deadline, or
\item[] (III) any job that has arrived, but with the constraint that
  if it is started, it must eventually be completed.
\end{description}

We consider all three metrics and all three versions of preemption. We
show that, for all three metrics, version~I is polynomially solvable,
and version~III is NP-complete. Many of the hardness results for no
preemption carry over to version~III. 

Our main results are for version~II. We show that the general problem
is NP-complete. Then, we focus on minimizing the makespan in two
complementary special cases: 
\begin{enumerate}

\item
All jobs have a common arrival time and arbitrary deadlines.

\item
 All jobs have a common deadline and arbitrary arrival times.  
\end{enumerate}

\noindent
We show that the first problem is
NP-complete, whereas the second problem can be solved in polynomial
time.

These last results illustrate a curious feature of the LBP. One can
convert one special case into the other by reversing the direction of
time. In the LBP, unlike many scheduling settings, this reversing of
time changes the complexity of the problem.

\begin{figure*}[bt] 
\def\RESULT#1{\vbox{\vskip 2pt\hbox{#1}\vskip 0.5pt}}
\begin{center}
\begin{tabular}{|l||c|c||c|}
\hline
{\bf Instance} & 
{\bf Preemption} & 
{\bf Metric} & 
{\bf Complexity}\\
\hline \hline
$\bullet $ Arbitrary deadlines  &
\RESULT{II} &
\RESULT{3} & 
\RESULT{(Weakly) NP-Complete} \\
$\bullet $ Identical job arrivals: &
\RESULT{~~~} & 
\RESULT{~~~} & 
\RESULT{Hard to approximate} \\
~~~~$a_1 =  \cdots = a_n = 0$ & 
\RESULT{~~~} & 
\RESULT{~~~} & 
\RESULT{~~~} \\
\hline 
$\bullet $ Arbitrary job arrivals &
\RESULT{II} &
\RESULT{3} & \RESULT{Polynomial time} \\
$\bullet $ Identical deadlines: & 
\RESULT{~~~} &
\RESULT{~~~} &
\RESULT{~~~} \\
~~~~$d_1 =  \cdots = d_n = D$  &
\RESULT{~~~} &
\RESULT{~~~} &
\RESULT{~~~} \\
\hline
\end{tabular}
\end{center}
\caption{
  Illustration that reversal of time changes complexity; see
  Section~\protect\ref{sec:pmtn-makespan}.}
\label{tab:time-reversal}
\end{figure*}

\subsection{Related Work on the LBP}

Recently Hepner and Stein~\cite{HepnerStein2002}
published a pseudo-polynomial-time algorithm
for minimizing the makespan subject to preemption constraint II,
thus resolving an open problem from
an earlier version of this paper~\cite{ArkinBenderMitchellSkiena99}.
They also extend the LBP to the parallel setting, 
in which there are multiple bureaucrats.

\section{LBP: No Preemption}
\label{nonpreempt}

In this section, we assume that no job can be preempted: if a job is
started, then it is performed without interruption until it completes.
We show that the Lazy Bureaucrat Problem (LBP) without preemption is
strongly NP-complete and is not approximable to within any factor for
the three metrics we consider.  These hardness results distinguish our
problem from traditional scheduling metrics, which can be approximated
in polynomial time, as proved in~\cite{BGNS-99}.  We show, however,
that several special cases of the problem have pseudo-polynomial-time
algorithms, using applications of dynamic programming.

\subsection{Hardness Results}
\label{sec:hard}

We begin by describing the relationship
between the three different objective functions 
from Section~\ref{sec:hard}
in the case of no preemption.  

The problem of minimizing the
total work (objective function 1) is a special case of the problem of 
minimizing the weighted sum of completed jobs 
(objective function 2), because without preemption 
every job that is
executed must be completed.  (The weights become the job lengths.)
Furthermore, if all jobs have
the same arrival time, say time zero, then the 
two objectives 
minimizing the total amount of time spent working
and minimizing  the makespan (go home early) are equivalent
(objective functions 1 and 3), since
no feasible schedule will have any gaps.
Our first hardness theorem
applies therefore to all three objective functions
from Section~\ref{sec:hard}.

\begin{thm}
\label{thm:hard1}
The Lazy Bureaucrat Problem with no preemption is (weakly) NP-complete
for objective functions (1)-(3), 
and is not approximable to within any fixed factor, even when all arrival
times are the same.
\end{thm}

\begin{pf}
We use a reduction from the {\sc Subset Sum} problem~\cite{GJ-79}:
Given a set of
integers $S=\{x_1,x_2,\ldots,x_n\}$ and a target integer $T$, does there 
exist a subset $S'\subseteq S$, such that $\sum_{x_i\in S'}x_i=T$?

We construct an instance of the LBP having $n+1$ jobs, each having
release time zero ($a_i=0$ for all $i$). For $i=1\ldots,n$, job $i$
has processing time $t_i=x_i$ and deadline $d_i=T$.  Job $n+1$ has
processing time $t_{n+1}=1+\sum_{x_i\in S}x_i$ and deadline
$d_{n+1}=T+t_{n+1}-1$; thus, job $n+1$ can be started at time $T-1$ or
earlier.  Because job $n+1$ is so long, the bureaucrat wants to avoid executing
it, but can do so if and
only if he selects a subset of jobs from $\{1,\ldots,n\}$ to execute
whose lengths sum to exactly~$T$.
In summary, the large job $n+1$ is executed if and only if
the subset problem is solved exactly and executing the long job
leads to a schedule whose makespan (i.e., total work executed)
is not within any fixed factor of the optimal solution.
\end{pf}

We now show that the LBP with no preemption
is strongly NP-complete.  
As we will show in Section~\ref{pseudo}, the LBP from Theorem~\ref{thm:hard1}
when all arrival times are equal,
has a pseudo-polynomial-time algorithm.
However, if arrival times and deadlines are arbitrary integers, the
problem becomes strongly NP-complete. 
Thus, the following theorem
subsumes Theorem~\ref{thm:hard1} when arrival times and deadlines
our unconstrained, whereas
Theorem~\ref{thm:hard1} is more generally applicable.

\begin{thm}
\label{thm:hard2}
The Lazy Bureaucrat Problem with no preemption is strongly
NP-complete for objective functions (1)--(3), and is not approximable to within any fixed factor.
\end{thm}

\begin{figure}[htbp]
\begin{center}
\input{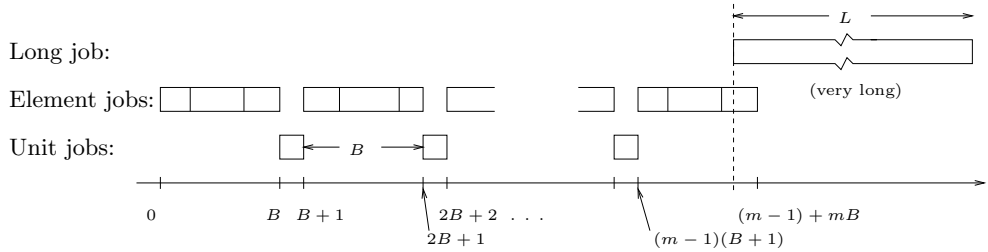}
\end{center}
\caption{
  Proof of hardness of LBP with no preemption and arbitrary arrival
  times.}
\label{fig:no-preempt-hard}
\end{figure}

\begin{pf}
Clearly the problem is in NP, since any solution can be represented
by an ordered list of jobs, given their arrival times.
To show hardness,
we use a reduction from the {\sc 3-Partition} problem~\cite{GJ-79}:
Given a
set $S=\{x_1,\ldots,x_{3m}\}$ of $3m$ positive integers and a positive
integer bound $B$ such that $B/4<x_i<B/2$, for $i=1,\ldots,3m$ and
$\sum_i x_i=mB$, does there exist a partitioning of $S$ into $m$
disjoint sets, $S_1,\ldots,S_m$, such that for $i=1,...,m$,
$\sum_{x_j\in S_i} x_j=B$? (Note that, by the assumption that
$B/4<x_i<B/2$, each set $S_i$ must contain exactly 3 elements.)

Objective function (1)
is a special case of
objective function (2)
because without preemption, any job that is begun must be completed.
Furthermore, hard instances will be designed so that there are no gaps,
ensuring that the optimal solution for objective function (1)
is also the optimal solution for objective function (3).

We construct an instance of the LBP containing three classes of jobs:

\begin{itemize}
\item {\em Element jobs\/} --- We define one ``element job'' corresponding to
each element $x_i\in S$, having arrival time~$0$, deadline
$d_i=(m-1)+mB$, and processing time~$x_i$.

\item {\em Unit jobs\/} --- 
 We define $m-1$ ``unit'' jobs, each of length
$1$. The $i$-th unit job (for $i=1,\ldots,m-1$) has arrival time
$i(B+1)-1$ and deadline $i(B+1)$. Note that for these unit-length jobs
we have $d_j-a_j=1$; thus, these jobs must be processed immediately
upon their arrival, or not at all.

\item {\em Large job\/} --- 
We define one ``large'' job of length $L>(m-1)+mB$,
arrival time $0$, and deadline $L+(m-2)+mB$. Note that in order to
complete this job, it must be started at time $(m-2)+mB$ or before.
\end{itemize}

As in the proof of Theorem~\ref{thm:hard1}, the lazy bureaucrat wants
to avoid executing the long job, but can do so if and only if all
other jobs are actually executed.  Otherwise, there will be a time
when the large job is the only job in the system and the lazy
bureaucrat will be forced to execute it.  Thus, the unit jobs must be
done immediately upon their arrival, and the element jobs must fit in
the intervals between the unit jobs.  Each such interval between
consecutive unit jobs is of length exactly $B$.  Refer to
Figure~\ref{fig:no-preempt-hard}. In summary, the long job is not
processed if and only if all of the element and unit jobs can be
processed before their deadlines, which happens if and only if the
corresponding instance of {\sc 3-Partition} is a ``yes'' instance.
Note that since $L$ can be as large as we want, this also implies that
no polynomial-time approximation algorithm with any fixed
approximation bound can exist, unless P=NP.  
\end{pf}

\subsection{Algorithms for Special Cases}
\label{pseudo}

\subsubsection{Unit-Length Jobs}

Consider the special case of the LBP in which all
jobs have unit processing times.
(Recall that all inputs are assumed to be integral.)
The Latest Due Date (LDD) scheduling policy 
selects the job in the system having the latest deadline.
Note that this policy in nonpreemptive for unit-length jobs, since
all jobs have integral arrival times.

\begin{thm}
Consider the Latest Deadline First scheduling policy 
when jobs have unit lengths and all inputs are integral.
The LDD scheduling policy
minimizes the amount of executed work.
\end{thm}

\begin{pf}
Assume by contradiction that no optimal schedule is LDD.  We use an
exchange argument. 
Consider an optimal (non-LDD) schedule that has
the fewest pairs of jobs executed in non-LDD order.  
The schedule must have two neighboring jobs $i,j$ such that $i<j$ in the
schedule but $D_i<D_j$, and $j$ is in the system when
$i$ starts its execution. Consider the first such pair of jobs. 
There are two cases:

(1) The new schedule with $i$ and $j$ switched, is feasible. It
executes no more work than the optimal schedules, and is therefore
also optimal.

(2) The schedule with $i$ and $j$ switched is not feasible. This
happens if $i$'s deadline has passed.  If no job is in the system
to replace $i$, then we obtain a better schedule than the optimal schedule
and reach a contradiction.
Otherwise, we replace $i$ with the other job and repeat the switching
process.

We obtain a schedule executing no more work than an optimal
schedule, but with fewer pairs of jobs in non-LDD order, a contradiction.
\end{pf}

\subsubsection{Narrow Windows}
Consider now the version in which jobs are large in comparison with
their intervals,  that is, the intervals are ``narrow.''
Let $R$ be a bound on the ratio of window length to job length; i.e.,
for each job $i$, $d_i-a_i< R \cdot t_i$. 
We show that a pseudo-polynomial algorithm exists for the case of
sufficiently narrow windows, that is,  when $R \leq 2$.

\begin{lem}
\label{lem:unique-ordering}
Assume that for each job $i$, $d_i-a_i<2t_i$.
Then, if job $i$ can be
scheduled before job $j$, then job $j$ cannot be scheduled before
job~$i$.
\end{lem}

\begin{pf}
We rewrite the assumption: for each $i$, $d_i-t_i< t_i+a_i$.
The fact that job $i$ can be scheduled before job $j$ is equivalent to
the statement that $a_i+t_i\leq d_j-t_j$, since the earliest that job
$i$ can be completed is at time $a_i+t_i$ and the latest that job $j$
can be started is at time $d_j-t_j$.  Combining these inequalities,
we obtain
$$a_j+t_j> d_j-t_j\geq a_i+t_i > d_i-t_i,$$
which implies that job $j$ cannot be scheduled before job~$i$.
\end{pf}

\begin{cor}
Under the assumption that $d_i-a_i<2t_i$ for each $i$, the ordering
of any subset of jobs in a schedule is uniquely determined.
\end{cor}

\begin{thm}
Suppose that for each job $i$, $d_i-a_i< 2 t_i$.
Let $K=\max_i d_i$.
Consider the problem of minimizing 
objective functions~(1)-(3) 
from Section~\ref{subsec-model}, 
in the nonpreemptive setting.
Then the LBP can be solved in $O(n K \max(n,K))$ time.
\label{fixed-window-DP}
\end{thm}

\begin{pf}
We use dynamic programming to find the shortest path in a 
directed acyclic graph (DAG).
There are $O(nK^2)$ states the system can enter.
Let $(i,j,\tau)$ denote the state of the system when the processor
begins executing the $j$-th unit of work of job $i$ at time $\tau$.
Thus, $i=1,...,n$, $j=1,...,t_i$, and $\tau =0,...,K$. Transitions
from state to state are defined according to the following rules:

\begin{enumerate}
\item No preemption: once a job is begun, it must be completed without
  interruptions.
\item When a job is completed at time $\tau$, another job must begin
  immediately if one exists in the system. (By
  Lemma~\ref{lem:unique-ordering}, we know this job has not yet been
  executed.) Otherwise, the system is idle and begins executing a job
  as soon as one arrives.
\item State $(i,t_i,\tau)$ is an end state if and only if when job $i$
  completes at time $\tau$, no jobs can be executed subsequently.
\item The start state has transitions to the jobs $(i,1,0)$ that
  arrive first.
\end{enumerate}

The goal of the dynamic program is to find the length of a shortest
path from the start state to an end state.  Depending on how we assign
weights to the edges we can force our algorithm to minimize all three
metrics from Section~\ref{subsec-model}.  To complete the time
analysis, note that only $n K$ of the $nK^2$ states have more than
constant outdegree, and these states each have outdegree bounded
by~$n$.
\end{pf}

For $R>2$ we know of no efficient algorithm without additional
conditions.
Let $W$ be a bound on the ratio of longest window to shortest window, and let
$\Delta$ be a bound on the ratio of the longest job to the shortest job.
Note that bounds on $R$ and $\Delta$ imply a bound on $W$, and
bounds on $R$ and $W$ imply a bound on $\Delta$.
However, a bound on $\Delta$ alone is not sufficient for a 
pseudo-polynomial-time algorithm.

\begin{thm}
\label{thm:bounded-Delta}
Even with a bound on the ratio $\Delta$, the LBP with no preemption
is strongly NP-complete for objective functions~(1)-(3).
It cannot be approximated to within  a factor of $\Delta - \epsilon$, 
for any $\epsilon>0$, unless P=NP.
\end{thm}

\begin{pf}
Modify the reduction from 3-partition of Theorem~\ref{thm:hard2},
by changing all the fixed ``unit'' jobs to have length $B/3$,
and adjust the arrival times and deadlines accordingly.

Instead of one very long job
as in the proof from Theorem~\ref{thm:hard2},
we create a sequence of bounded-length jobs that serve the same purpose.
One unit before the deadline of the ``element'' jobs
(see Theorem~\ref{thm:hard2}) a sequence of longer jobs
$\ell_1,\ldots,\ell_m$ arrives.  Each job $\ell_i$ entirely fills its window
and so can only be executed directly when it arrives.  Job  $\ell_{i+1}$
arrives at the deadline of job  $\ell_i$.  In addition, a sequence of
shorter jobs $s_1,\ldots,s_m$ arrives, where each
shorter job $s_i$ also entirely fills its window and can only be executed 
when it arrives.  Shorter job $s_i$ overlaps $\ell_i$ and $\ell_{i+1}$; it
arrives one unit before the deadline of
job $\ell_i$. 
Jobs $s_i$ have length $B/4$ and jobs $\ell_i$ have length $\Delta\cdot B/4$.
 Thus, if all the jobs comprising the
$3$-partition problem can be executed, jobs $\ell_1,\ldots,\ell_m$
will be avoided by executing jobs $s_1,\ldots,s_m$.  Otherwise, jobs
$\ell_1,\ldots,\ell_m$ must be executed.  The index of $m$ can be adjusted
to any $\epsilon$.
\end{pf}

Bounds on both $\Delta$ and $R$ are sufficient to yield a
pseudo-polynomial algorithm:

\begin{thm}
Let $K = \max_i d_i$.
Given bounds on $R$ and $\Delta$,
the Lazy Bureaucrat Problem with no preemption
can be solved in $O(K\cdot n^{4 R\lg \Delta})$
for objective functions~(1)-(3).
\end{thm}

\begin{pf}
We modify the
dynamic programming algorithm of Theorem~\ref{fixed-window-DP}
for this more complex situation.
The set of jobs potentially available to work on in a given schedule
at time $\tau$ are the jobs $j$ that have not yet been executed, for which 
$d_j - t_j \geq \tau$.
Our state space will encode the {\em complement\/} of this set for each time 
$\tau$,
specifically, the set of jobs that were executed earlier but
could otherwise have been executed at time $\tau$.

The bounds on $R$ and $\Delta$ together imply an upper bound on 
the number of subsets of jobs active at time 
$\tau$ that could have been executed
prior to time $\tau$.
Let $d_{\min}$ be the length of the shortest job potentially 
active at time $\tau$.
We can partition all potentially active jobs into $\lg \Delta$ classes,
where the $j$th class consists of the jobs of size
greater than or equal to $2^{j-1} d_{\min}$
and less than  $2^j d_{\min}$.
The earliest possible arrival time of any class-$j$ job
is $\tau - 2^j d_{\min} R$, since each job has an $R$-bounded window.
Only $2^j d_{\min} R / 2^{j-1} d_{\min} = 2R$ jobs from class $j$
can be executed within this window.
Summing over all the classes implies that at most $2 R \lg \Delta$ jobs
potentially active at time $\tau$ could have been executed in a non-preemptive
schedule by time $\tau$.
As before the choice of weights on the edges
determines the metric that is optimized.

The time bound on the running time 
follows by observing that each of the $K n^{(2 R \lg \Delta)}$
states has outdegree at most $2 R \lg \Delta$.
\end{pf}

\subsubsection{Jobs Having a Common Release Time}

In the next version of the problem all jobs are released at time zero,
i.e., $a_i=0$ for all $i$.  This problem can be solved in
pseudo-polynomial time by dynamic programming, specifically, reducing
the problem to that of finding the shortest path in a directed acyclic
graph.  The dynamic programming works because of the following
structural result: There exists an optimal schedule that executes the
jobs Earliest Due Date (EDD).

In fact this problem is a special case of the following general
problem: Minimizing the weighted sum of jobs not completed by their
deadlines.  A similar problem was solved by~\cite{LawlerMoore69}, using the
same structural result.

\begin{thm}
The LBP can be solved in pseudo-polynomial time 
for all three metrics when all jobs have
a common release time.
Specifically, let $K = \max_i d_i$; then the running time is
$O(K\cdot n).$
\end{thm}

\section{LBP: Preemption}
\label{sec:preempt}

In this section we consider the Lazy Bureaucrat Problem in which 
jobs may be preempted: a job in progress can be set aside, while another
job is processed, and then possibly resumed later.  It is important to
distinguish among different constraints that specify which jobs
are available to be processed.  We consider three natural choices of such 
constraints:

\begin{description}
\item[Constraint I:] In order to work on job $i$ at time $\tau$, we
require only that the current time $\tau$ lies within the job's
interval $I_i$: $a_i\leq \tau\leq d_i$.

\item[Constraint II:] In order to work on job $i$ at time $\tau$, we
require not only that the current time $\tau$ lies within the job's
interval $I_i$, but also that the job has a {\em chance\/} to be
completed, e.g., if it is processed without interruption until
completion.

This condition is equivalent to requiring that $\tau\leq c'_i$, where
$c'_i=d_i-t_i+y_i$ is the {\em adjusted critical time\/} of job $i$:
$c'_i$ is the latest possible time to start job $i$, in order to meet
its deadline $d_i$, given that an amount $y_i$ of the job has already
been completed.

\item[Constraint III:] In order to work on job $i$, we require that
$\tau\in I_i$.  Further, we require that any job that is started
is eventually completed.

\end{description}

We divide this section into subsections,
where each subsection considers one of 
the three objective functions (1)--(3)
from Section~\ref{subsec-model}, in
which the goals are to minimize (1) the total time working (regardless of
which jobs are completed), (2) the weighted sum of completed jobs, or
(3) the makespan of the schedule (the ``go home'' time).
For each metric we see that the constraints on preemption
can dramatically affect the complexity of the problem.

Constraint III makes the LBP with preemption quite similar
to the LBP with no preemption. In fact, if all jobs arrive at the same
time ($a_i=0$ for all $i$), then the three objective functions are
equivalent, and the problem is hard:

\begin{thm}
\label{hard-pre-3}
The LBP with preemption, under constraint~III (one must complete any
job that is begun), is (weakly)
NP-complete and hard to approximate for all three objective
functions.
\end{thm}

\begin{pf}
We use the same reduction as the one given in the proof of
Theorem~\ref{thm:hard1}.  Note that any schedule for an instance given
by the reduction, in which all jobs processed must be completed
eventually, can be transformed into an equivalent schedule with no
preemptions. This makes the problem of finding an optimal schedule
with no preemption equivalent to the problem of finding an optimal
schedule in the preemptive case under constraint III.  
\end{pf}

Note that we
cannot use a proof similar to that of Theorem~\ref{thm:hard2} to show
that this problem is strongly NP-complete, since preemption can lead
to improved schedules in that instance.

\subsection{Minimizing Total Time Working}

\begin{thm}
\label{thm:preempt-I.1}
The LBP with preemption, under constraint~I (one can work on any job
in its interval) and objective~(1) (minimize total time
working), is polynomially solvable.
\end{thm}

\begin{pf}
The algorithm schedules jobs according to latest due date
(LDD), in which at all times the job in the system with the latest
deadline is being processed, with ties broken arbitrarily.  An
exchange argument shows that this is optimal.  Suppose there is an
optimal schedule that is not LDD. Consider the first time in which an
optimal schedule differs from LDD, and let OPT be an optimal schedule
in which this time is as late as possible. Let OPT be executing a
piece of job $i$, $p_i$ and LDD executes a piece of job $j$, $p_j$. We
know that $d_i<d_j$. We want to show that we can replace the first
unit of $p_i$ by one unit of $p_j$, contradicting the choice of OPT,
and thereby proving the claim. If in OPT, job $j$ is not completely
processed, then this swap is feasible, and we are done. On the other
hand, if all of $j$ is processed in OPT,  such a swap causes a unit of
job $j$ later on to be removed, leaving a gap of one unit. If this gap
cannot be filled by any other job piece, we get a schedule with less
work than OPT, which is a contradiction. Therefore assume the gap can
be filled, possibly causing a later unit gap. Continue this process,
and at its conclusion, either a unit gap remains contradicting the
optimality of OPT, or no gaps remain, contradicting the choice of OPT.
\end{pf}

\begin{thm}
The LBP with preemption, under constraint~II (one can only work on
jobs that can be completed) and objective~(1) (minimize total time
working), is (weakly) NP-complete.
\end{thm}

\begin{pf}
If all arrival times are the same, then this problem is equivalent to
the one in which the objective function is to minimize the makespan,
which is shown to be NP-complete in Theorem~\ref{thm:preempt-II.3}. 
\end{pf}

\subsection{Minimizing Weighted Sum of Completed Jobs}

\begin{thm}
The LBP with preemption, under constraint~I (one can work on any job
in its interval) and objective~(2) (minimize the weighted sum of
completed jobs), is polynomially solvable.
\end{thm}

\begin{pf}
Without loss of generality, assume that jobs $1,\ldots,n$ are
indexed in order of increasing deadlines. We show how to decompose the
jobs into separate components that can be treated independently.
Schedule the jobs according to EDD (if a job is executing and its
deadline passes, preempt and execute the next job).  Whenever there is
a gap (potentially of size zero), where no jobs are in the system, the
jobs are divided into separate components that can be scheduled
independently and their weights summed.  Now we focus on one such set
of jobs (having no gaps).  We modify the EDD schedule by preempting a
job $\epsilon$ units of time before it completes.  Then we move the
rest of the jobs of the schedule forward by $\epsilon$ time units and
continue to process.  At the end of the schedule, there are two
possibilities.  (1) the last job is interrupted because its deadline
passes; in this case we obtain a schedule in which no jobs are
completed; (2) the last job completes and in addition
all other jobs whose deadlines have not passed are also forced to
complete. 

The proof is completed by noting the following:
\begin{itemize}
\item There is an optimal schedule that completes all of its jobs at
  the end; and
\item The above schedule executes the maximum amount of work possible.
  (In other words, EDD (``minus $\epsilon$'') allows one to execute
  the maximum amount of work on jobs 1 through $i$ without completing
  any of them.)
\end{itemize}
\end{pf}

\begin{thm}
The LBP with preemption, under constraint~II (one can only work on
jobs that can be completed) and objective~(2) (minimize the weighted
sum of completed jobs), is (weakly) NP-complete.
\end{thm}

\begin{pf}
  Consider the LBP under constraint~II, where the objective is to
  minimize the makespan.  The proof of Theorem~\ref{thm:preempt-II.3}
  will have hard instances where all jobs have the same arrival time,
  and where the optimal solution completes any job that it begins.
  Thus, for these instances the metric of minimizing the makespan is
  equivalent to the metric of minimizing the weighted sum of completed
  jobs, for weights proportional to the processing times.
\end{pf}

\subsection{Minimizing Makespan: Going Home Early}
\label{sec:pmtn-makespan}

We assume now that the bureaucrat's goal is to go home as soon as
possible.

We begin by noting that if the arrival times are all the same
($a_i=0$, for all $i$), then the objective (3) (go home as soon as
possible) is in fact equivalent to the objective (1) (minimize total
time working), since, under any of the three constraints I--III, the
bureaucrat will be busy nonstop until he can go home.

Observe that if the {\em deadlines\/} are all the same
($d_i=D$, for all $i$), then the objectives (1) and (3) are quite
different.  Consider the following example.  Job~1 arrives at time
$a_1=0$ and is of length $t_1=2$, job~2 arrives at time $a_2=0$ and is
of length $t_2=9$, job~3 arrives at time $a_3=8$ and is of length
$t_3=2$, and all jobs have deadline $d_1=d_2=d_3=10$.  Then, in order
to minimize total time working, the bureaucrat will do jobs 1 and 3, a
total of 4 units of work, and will go home at time 10.  However, in
order to go home as soon as possible, the bureaucrat will do job~2,
performing 9 units of work, and go home at time 9 (since there is not
enough time to do either job~1 or job~3).

\begin{thm}
\label{thm:preempt-I.3}
The LBP with preemption, under constraint~I (one can do any job in its
interval) and objective~(3) (go home as early as possible), is
polynomially solvable.
\end{thm}

\begin{pf}
The algorithm is to schedule by Latest Due Date (LDD). 
The proof is similar to the one given in Theorem~\ref{thm:preempt-I.1}.
\end{pf}

If instead of constraint~I we impose constraint~II, the problem becomes hard:

\begin{thm}
\label{thm:preempt-II.3}
The LBP with preemption, under constraint~II (one can only work on jobs that
can be completed) and objective~(3) (go home as early as possible), is
(weakly) NP-complete, even if all arrival times are the same.
\end{thm}

\begin{pf}
 We give a reduction from {\sc Subset Sum}.  Consider an
instance of {\sc Subset Sum} given by a set $S$ of $n$ positive integers,
$x_1$, $x_2,\ldots,x_n$, and target sum $T$.  We construct an instance
of the required version of the LBP as
follows.  For each integer $x_i$, we have a job $i$ that arrives at
time $a_i=0$, has length $t_i=x_i$, and is due at time
$d_i=T+x_i-\epsilon$, where $\epsilon$ is a small constant (it
suffices to use $\epsilon={1\over 3n}$).  In addition, we have a
``long'' job $n+1$, with length $t_{n+1}>T$, that arrives at
time $a_{n+1}=0$ and is due at time $d_{n+1}=T-2\epsilon+t_{n+1}$.
We claim that it is possible for the bureaucrat to go home by time $T$
if and only if there exists a subset of $\{x_1,\ldots,x_n\}$ that sums to
exactly~$T$.

If there is a subset of $\{x_1,\ldots,x_n\}$ that sums to exactly~$T$,
then the bureaucrat can perform the corresponding subset of jobs (of
total length $T$) and go home at time $T$; he is able to avoid doing
any of the other jobs, since their critical times fall at an earlier
time ($T-\epsilon$ or $T-2\epsilon$), making it infeasible to begin
them at time $T$, by our assumption.

If, on the other hand, the bureaucrat is able to go home at time $T$,
then we know the following:

\begin{enumerate}
\item {\em The bureaucrat must have just completed a job at time
    $T$.\/}
  
  He cannot quit a job and go home in the middle of a job, since the
  job must have been completable at the instant he started (or
  restarted) working on it, and it remains completable at the moment
  that he would like to quit and go home.

\item {\em The bureaucrat must have been busy the entire time from 0
    until time $T$.\/}
  
  He is not allowed to be idle for any period of time, since he could
  always have been working on some available job, e.g., job $J_{n+1}$.
  
\item {\em If the bureaucrat starts a job, then he must finish it.\/}
  
  First, we note that if he starts job $J_i$ and does at least
  $\epsilon$ of it, then he must finish it, since at time $T$ less
  than $x_i-\epsilon$ remains to be done of the job, and it is not due
  until time $T-\epsilon+x_i$, making it feasible to return to the job
  at time $T$ (so that he cannot go home at time $T$).
  
  Second, we must consider the possibility that he may perform very
  small amounts (less than $\epsilon$) of some jobs without finishing
  them.  However, in this case, the {\em total\/} amount that he
  completes of these barely started jobs is at most $n\epsilon\leq
  {1\over 3}$.  This is a contradiction, since his total work time
  consists of this fractional length of time, plus the sum of the
  integral lengths of the jobs that he completed, which cannot add up
  to the integer $T$.  Thus, in order for him to go home at exactly
  time $T$, he must have completed every job that he started.
  
  Finally, note that he cannot use job $J_{n+1}$ as ``filler'', and do
  part of it before going home at time $T$, since, if he starts it and
  works at least time $2\epsilon$ on it, then, by the same reasoning
  as above, he will be forced to stay and complete it.  Thus, he will
  not start it at all, since he cannot complete it before time $T$
  (recall that $t_{n+1}>T$).

\end{enumerate}
We conclude that the bureaucrat must complete a set of jobs whose lengths
sum exactly to~$T$.

Thus, we have reduced {\sc Subset Sum} to our problem, showing that it is
(weakly) NP-complete.

Note that the LBP we have constructed has non integer data. However,
we can ``stretch'' time to get an equivalent problem in which all the
data is integral. Letting $\epsilon = \frac{1}{3n}$, we multiply all job
lengths and due dates by~$3n$.

\end{pf}

\begin{figure}[htbp]
\begin{center}
\input{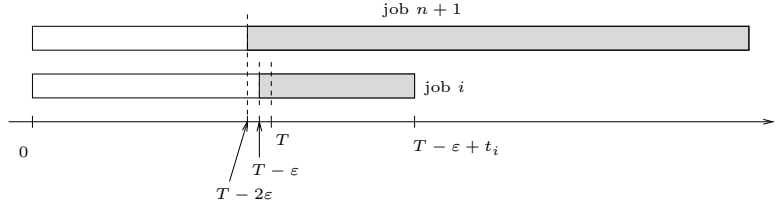}
\end{center}
\caption{
  Proof of hardness of LBP with preemption, assuming that all arrival
  times are at time 0.}
\label{fig:preempt-hard}
\end{figure}

\noindent {\em Remark.}\quad 
Hepner and Stein~\cite{HepnerStein2002} recently published a
pseudo-polynomial-time algorithm for this problem, thus resolving an
open problem from an earlier version of this
paper~\cite{ArkinBenderMitchellSkiena99}.  
\medskip

We come now to one of the main results of the paper.  We emphasize
this result because it uses a rather sophisticated algorithm and
analysis in order to show that, in contrast with the case of identical
arrival times, the LBP with identical deadlines is polynomially
solvable.  Specifically, the problems addressed in
Theorems~\ref{thm:preempt-II.3}~and~\ref{thm:preempt-II.3-same-dead}
are identical except that the flow of time is reversed.  Thus, we
demonstrate that in contrast to most classical scheduling problems, in
the LBP, when time flows in one direction, the problem is NP-hard,
whereas when the flow of time is reversed, the problem is
polynomial-time solvable.

The remainder of this section is devoted to proving the following
theorem:

\begin{thm}
\label{thm:preempt-II.3-same-dead}
The LBP with preemption, under constraint~II (one can only work on
jobs that can be completed) and objective~(3) (go home as early as
possible), is solvable in polynomial time if all jobs have the same
deadlines ($d_i=D$, for all $i$).  
\end{thm}

We begin with a definition a ``forced gap:'' There is a {\em forced
  gap\/} starting at time $\tau$ if $\tau$ is the earliest time such
that the total work arriving by time $\tau$ is less than $\tau$.  This
(first) forced gap ends at the arrival time, $\tau'$, of the next job.
Subsequently, there may be more forced gaps, each determined by
considering the scheduling problem that starts at the end, $\tau'$, of
the previous forced gap.  We note that a forced gap can have length
zero.

Under the ``go home early'' objective, we can assume, without loss of
generality, that there are no forced gaps, since our problem really
begins only at the time $\tau'$ that the {\em last\/} forced gap ends.
(The bureaucrat is certainly not allowed to go home before the end
$\tau'$ of the last forced gap, since more jobs arrive after $\tau'$
that can be processed before their deadlines.)  While an optimal
schedule may contain gaps that are not forced, the next lemma implies
that there exists an optimal schedule having no unforced gaps.

\begin{lem}
  Consider the LBP of Theorem~\ref{thm:preempt-II.3-same-dead}, and
  assume that there are no forced gaps.  If there is a schedule having
  makespan $T$, then there is a schedule with no gaps, also having
  makespan~$T$.
\end{lem}

\begin{pf}
  Consider the first gap in the schedule, which begins at time $g$.
  Because the gap is not forced, there is some job $j$ that is not
  completed, and whose critical time is at time $g'\leq g$.  This is
  because there must be a job that arrived before $g$ that is not
  completed in the schedule, and at time $g$ it is no longer feasible
  to complete it, and therefore its critical time is before $g$.  The
  interval of time between $g'$ and $T$ may consist of (1) gaps, (2)
  work on completed jobs, and (3) work on jobs that are never
  completed. Consider a revised schedule in which, after time $g'$,
  jobs of type~3 are removed, and jobs of type~2 are deferred to the
  end of the schedule. (Since a job of type~2 is completed and all
  jobs have the same deadline, we know that it is possible to move it
  later in the schedule without passing its critical time.  It may not
  be possible to move a (piece of a) job of type~3 later in the
  schedule, since its critical time may have passed.)  In the revised
  schedule, extend job~$j$ to fill the empty space. Note that there is
  enough work in job $j$ to fill the space, since a critical time of
  $g'$ means that the job must be executed continuously until deadline
  $D$ in order to complete it.
\end{pf}

\begin{lem} 
  Consider an LBP of Theorem~\ref{thm:preempt-II.3-same-dead} in which
  there are no forced gaps.  Any feasible schedule can be rearranged
  so that all completed jobs are ordered by their arrival times and
  all incomplete jobs are ordered by their arrival times.
\end{lem}

\begin{pf}
  The proof uses a simple exchange argument as in the standard proof
  of optimality for the EDD (Earliest Due Date) policy in traditional
  scheduling problems.
\end{pf}

Our algorithm checks if there exists a schedule having no gaps that
completes exactly at time $T$. Assume that the jobs $1,\ldots,n$ are
labeled so that $a_1\leq a_2\cdots\leq a_n$.  The main steps of the
algorithm are as follows:

\subsection*{The Algorithm:}

\begin{enumerate}
  
\item Determine the forced gaps.  This allows us to reduce to a
  problem having no forced gaps, which starts at the end of the last
  forced gap.
  
  The forced gaps are readily determined by computing the partial
  sums, $\tau_j=\sum_{i=1}^j t_i$, for $j=0,1,\ldots,n$, and comparing
  them to the arrival times.  (We define $\tau_0=0$.) The first forced
  gap, then, begins at the time $\tau=\tau_{j^*}=\min\{\tau_j: \tau_j
  < a_{j+1}\}$ and ends at time $a_{j^*+1}$.  ($\tau=0$ if $a_1>0$;
  $\tau=\infty$ if there are no forced gaps.)  Subsequent forced gaps,
  if any, are computed similarly, just by re-zeroing time at~$\tau'$,
  and proceeding as with the first forced gap.
  
\item Let $x=D-T$ be the length of time between the common deadline
  $D$ and our target makespan $T$.  A job $i$ for which $t_i\leq x$ is
  called {\em short\/}; jobs for which $t_i> x$ are called {\em
    long\/}.
  
  If it is {\em not\/} possible to schedule the set of short jobs so
  that each is completed and they are all done by time $T$, then our
  algorithm stops and returns ``NO,'' concluding that going home by
  time $T$ is impossible.  Otherwise, we continue with the next step
  of the algorithm.
  
  The rationale for this step is the observation that any job of
  length at most $x$ must be completed in any schedule that permits
  the bureaucrat to go home by time $T$, since its critical time
  occurs at or after time~$T$.
  
\item Create a schedule ${\mathcal S}$ of all of the jobs, ordered by
  their arrival times, in which the amount of time spent on job $i$ is
  $t_i$ if the job is short (so it is done completely) and is $t_i-x$
  if the job is long.
  
  For a long job $i$, $t_i-x$ is the maximum amount of time that can
  be spent on this job without committing the bureaucrat to completing
  the job, i.e., without causing the adjusted critical time of the job
  to occur after time~$T$.
  
  If this schedule ${\mathcal S}$ has no gaps and ends at a time after
  $T$, then our algorithm stops and returns ``YES.'' A feasible
  schedule that allows the bureaucrat to go home by time $T$ is
  readily constructed by ``squishing'' the schedule that we just
  constructed: We reduce the amount of time spent on the long jobs,
  starting with the latest long jobs and working backwards in time,
  until the completion time of the last short job exactly equals~$T$.
  This schedule completes all short jobs (as it should), and does
  partial work on long jobs, leaving all of them with adjusted
  critical times that fall {\em before\/} time $T$ (and are therefore
  not possible to resume at time $T$, so they can be avoided).
  
\item If the above schedule ${\mathcal S}$ has gaps or ends before
  time $T$, then ${\mathcal S}$ is not a feasible schedule for the
  lazy bureaucrat, so we must continue the algorithm.
  
  Our objective is to decide {\em which\/} long jobs to complete, that
  is, is there a set of long jobs to complete that will make it
  possible to go home by time~$T$.  This problem is solved using the
  dynamic programming algorithm {\em Schedule-by-$T$\/}, which is
  described in detail below.
\end{enumerate}

\subsection*{Procedure Schedule-by-{$T$}:}

Let $G_i$ be the sum of the gap lengths that occur before time $a_i$
in schedule ${\mathcal S}$. Then, we know that in order to construct a
gapless schedule, at least $\lceil G_i/x\rceil$ long jobs (in addition
to the short jobs) from $1,\ldots,i-1$ must be completed. For each $i$
we have such a constraint; collectively, we call these the {\em gap
  constraints\/}.

\begin{claim}
  If for each gap in schedule ${\mathcal S}$, there are enough long
  jobs to be completed in order to fill the gap, then a feasible
  schedule ending at $T$ exists.
\end{claim}

We devise a dynamic programming algorithm as follows.  Let $T(m,k)$ be
the earliest completion time of a schedule that satisfies the
following:
\begin{enumerate}
\item
It completes by time $T$;

\item
It uses jobs from the set $\{1,\ldots,k\}$;

\item
It completes exactly $m$ jobs and does no other work (so it
may have gaps, making it an infeasible schedule);

\item
It satisfies the gap constraints; and

\item
It completes all short jobs (of size $\leq x$).

\end{enumerate}
The boundary conditions on $T(m,k)$ are given by:
\begin{description}

\item[] $T(0,0)=0$; 

\item[] $T(0,n)=\infty$, which implies that at least one of the jobs
must be completed;

\item[] $T(m,0)=\infty$ for $m>0$;

\item[] $T(m,k)=\infty$ if there exist constraints such that at least
$m+1$ jobs from $1,\ldots,k$ must be completed, some of the jobs from
$1,\ldots,k$ must be completed because they are short, and some
additional jobs may need to be completed because of the gap
constraints.  Note that this implies that $T(0,k)$ is equal to zero or
infinity, depending on whether gap constraints are disobeyed.

\end{description}

In general, $T(m,k)$ is given by selecting the better of two options:
$$T(m,k)=\min\{\alpha ,\beta\},$$
where $\alpha$ is the earliest completion time if we choose not to
execute job $k$ (which is a legal option only if job $k$ is long),
giving
$$\alpha= \cases{T(m,k-1) & if  $t_k>x$ \cr
\infty & otherwise,}$$
and $\beta$ is the earliest completion time if we choose to execute
job $k$ (which is a legal option only if the resulting completion time
is by time $T$), giving
$$\beta= \cases{\max (a_k+t_k,T(m-1,k-1)+t_k) & if this quantity is $\leq T$  \cr
\infty & otherwise.}$$

\begin{lem}
  There exists a feasible schedule completing at time $T$ if and only
  if there exists an $m$ for which $T(m,n)<\infty$.
\end{lem}

\begin{pf}
  If $T(m,n)=\infty$ for all $m$, then, since the gap constraints
  apply to any feasible schedule, and it is not possible to find such
  a schedule for any number of jobs $m$, there is no feasible schedule
  that completes on or before~$T$.
  
  If there exists an $m$ for which $T(m,n)<\infty$, let $m^*$ be the
  smallest such $m$.  Then, by definition, $T(m^*,n)\leq T$. We show
  that the schedule ${\mathcal S^*}$ obtained by the dynamic program
  can be made into a feasible schedule ending at $T$. Consider jobs
  that are not completed in the schedule ${\mathcal S}^*$; we wish to
  use some of them to ``fill in'' the schedule to make it feasible, as
  follows.
  
  Ordered by arrival times of incomplete jobs, and doing up to $t_i-x$
  of each incomplete job, fill in the gaps. Note that by the gap
  constraints, there is enough work to fill in all gaps. There are two
  things that may make this schedule infeasible: (i) Some jobs are
  worked on beyond their critical times, and (ii) the last job to be
  done must be a completed one.
  
  (i). {\em Fixing the critical time problem:}\quad Consider a job $i$
  that is processed at some time, beginning at $\tau$, after its
  critical time, $c_i$.  We move all completed job pieces that fall
  between $c_i$ and $T$ to the end of the schedule, lining them up to
  end at $T$; then, we do job $i$ from time $c_i$ up until this batch
  of completed jobs. This is legal because all completed jobs can be
  pushed to the end of the schedule, and job $i$ cannot complete once
  it stops processing.
  
  (ii). {\em Fixing the last job to be a complete one:}\quad Move a
  ``sliver'' of the last completed job to just before time $T$. If
  this is not possible (because the job would have to be done before
  it arrives), then it means that we {\em must\/} complete one
  additional job, so we consider $m^*+1$, and repeat the process.
  
  Note that a technical difficulty arises in the case in which the sum
  of the gap lengths is an exact multiple of $x$: Do we have to
  complete an additional job or not? This depends on whether we can
  put a sliver of a completed job at $T$. There are several ways to
  deal with this issue, including conditioning on the last completed
  job, modifying the gap constraint, or ignoring the problem and
  fixing it if it occurs (that is if we cannot put a sliver, then add
  another job which must be completed to the gap constraint).
\end{pf}

This completes the proof of our main theorem,
Theorem~\ref{thm:preempt-II.3-same-dead}.

\noindent {\em Remark.}\quad Even if all arrival times are the same,
and deadlines are the same, and data is integer, an optimal solution
may not be integer. In fact, there may not be an optimal solution,
only a limiting one, as the following example shows: Let $a_i=0$ and
$d_i=100$, for all $i$.  Jobs $1,\ldots,n-1$ have length 51, while job
$n$ has length $t_{n}=48$.  A feasible schedule executes $\epsilon$ of
each of the first $n-1 $ jobs, where $\epsilon >1/(n-2)$, and all of
job~$n$, so the total work done is $(n -1) \epsilon
+48>(n-1)/(n-2)+48$. Note that each of the first $n-1$ jobs have
$51-\epsilon$ remaining to do, while there is only time
$52-(n-1)\epsilon$ left before the deadline.  Now, by making
$\epsilon$ arbitrarily close to $1/(n-2)$, we can make the schedule
better and better.

\section*{Acknowledgements}

We thank the referees for constructive comments and suggestions
that improved the presentation of the paper.


\begin{thebibliography}{00}

\bibitem{ArkinBenderMitchellSkiena99}
E.~M. Arkin, M.~A. Bender, J.~S.~B. Mitchell, and S.~S. Skiena.
\newblock The Lazy Bureaucrat Scheduling Problem.
\newblock {\em Proc.~6th Workshop on Discrete Algorithms (WADS)},
  pages 80--85, 1999.

\bibitem{BGNS-99}
A.~Bar-Noy, S.~Guha, J.~Naor, and B.~Schieber.
\newblock Approximating the throughput of real-time multiple machine
  scheduling.
\newblock In {\em Proc.~31st ACM Symp. Theory of Computing (STOC)}, 
pages 622--631, 1999.

\bibitem{BarvinokJWW1998}
A.~ I.~Barvinok, D.~S.~Johnson, G.~J.~Woeginger, and R.~Woodroofe.
\newblock The maximum traveling salesman problem under
polyhedral norms. 
\newblock In
{\em Proc.~6th Conference on
Integer Programming and Combinatorial Optimization (IPCO)\/},
{\em Lecture Notes in Computer Science}, vol.~1412, 
pages 195--201, 1998.

\bibitem{Fekete1999}
S.~P.~Fekete.
\newblock Simplicity and hardness of the maximum traveling 
salesman problem under geometric distances. 
\newblock {\em Proc.~10th {ACM-SIAM} 
Symposium on Discrete Algorithms (SODA)}, pages
337--345, 1999.

\bibitem{GJ-79}
M.~R. Garey and D.~S. Johnson.
\newblock {\em Computers and Intractability: A Guide to the Theory of
  {NP}-Completeness}.
\newblock W. H. Freeman, San Francisco, 1979.

\bibitem{GoemansWilliamson1995}
M.~X.~Goemans and D.~P.~Williamson.
\newblock Improved approximation algorithms for maximum
cut and satisfiability problems using semidefinite programming.
\newblock {\em J. ACM}, 42:1115--1145, 1995.

\bibitem{HassinRubinstein1998}
R.~Hassin and S.~Rubinstein. 
\newblock An approximation algorithm for the maximum
traveling salesman problem. 
\newblock {\em Information Processing Letters}, 67(3):125-130, 1998.

\bibitem{HepnerStein2002}
C.~Hepner and C.~Stein.
\newblock Minimizing makespan for the lazy bureaucrat problem.
\newblock In  {\em Proc.~8th
Scandinavian Workshop on Algorithm Theory (SWAT)},
pages 40--50, 2002.

\bibitem{KargerMotwaniRamkumar1997}
D. Karger, R. Motwani, and G. Ramkumar. 
\newblock On approximating the longest path in a graph. 
\newblock {\em Algorithmica}, 18(1):82-98, 1997.

\bibitem{KargerSteinWein:1997}
D.~Karger, C.~Stein, and J.~Wein.
\newblock Scheduling algorithms.
\newblock 
In {\em Algorithms and Theory of Computation Handbook}, 
CRC Press, 1999.

\bibitem{Keneally-82}
T.~Keneally.
\newblock {\em Schindler's List}.
\newblock Touchstone Publishers, New York, 1993.

\bibitem{KosarajuParkStein1994}
S.~R.~Kosaraju, J.~ K.~Park, and C.~Stein. 
\newblock Long tours and short superstrings (preliminary version). 
\newblock In {\em Proc.~35th Annual Symposium on Foundations of
Computer Science (FOCS)}, pages 166--177, 1994. 

\bibitem{LLKS-93}
E.~L.~Lawler, J.~K. Lenstra, A.~H.~G. Rinnooy Kan, and D.~B.~Shmoys.
\newblock Sequencing and scheduling: Algorithms and complexity.
\newblock S.C. Graves, P.H. Zipkin, and A.H.G. Rinnooy Kan (eds.) 
\newblock In {\em Logistics of
 Production and Inventory: Handbooks in Operations Research and Management Science},
  volume~4, pages 445--522.  North-Holland, Amsterdam, 445-522, 1993.

\bibitem{LawlerMoore69}
E.~L. Lawler and J.~M. Moore.
\newblock A functional equation and its application to resource allocation and
  sequencing problems.
\newblock {\em Management Science}, 16:77--84, 1969.

\bibitem{Pinedo-95}
M.~Pinedo.
\newblock {\em Scheduling: {Theory}, Algorithms, and Systems}.
\newblock Prentice Hall, 1995.

\end{thebibliography}
\end{document}